# E-COMMERCE FOR RURAL MICRO-ENTREPRENEURS: MAPPING RESTRICTIONS, ECOLOGIES OF USE AND TRENDS FOR DEVELOPMENT

Aditi Bhatia-Kalluri, Faculty of Information, University of Toronto, aditi.bhatia@mail.utoronto.ca

**Abstract:** This paper addresses the struggle of rural micro-entrepreneurs in the Global South in utilizing e-commerce to reach wider markets. This research paper looks at the adoption of e-commerce as a sustainable marketplace by the micro-entrepreneur sellers from the lower socio-economic rural communities in India, a booming digital economy in the Global South. 'Sustainability' here refers to a model for sustainable economic development sustaining the e-commerce as business model for the rural micro-entrepreneurs to flourish. This paper explores rural development by dismantling the factors that shape the ways technology and trade impact micro-entrepreneurs. The aim is to offer recommendations and solutions to contribute building the e-commerce as a sustainable marketplace for rural micro-entrepreneurs. Recent information and economic policy changes in India, along with the expansion of mobile infrastructure and a growing user base in rural regions makes this research timely and important. By scrutinizing the infrastructure and auditing the information needs and challenges of users, this research will illuminate the gaps that are leading to a lack of sustainable economic development, and information asymmetries discouraging the rural micro-entrepreneurs from selling online. The purpose of the paper is to find hurdles in the sustainable development of e-commerce as a business solution.

## 1. INTRODUCTION

### 1.1 Significance of the Study and Background

Despite the current availability of technology, networks, and infrastructure, microentrepreneurs in rural India have yet to capitalize on e-commerce. According to Kalambe (2019), "rural India accounts for two-thirds of the country's population" (p. 29) and their active participation in e-commerce can contribute to the exponential growth of e-commerce. This growth has to come with sustainability or sustainable development, which means longevity of information technology. Sustainable economic development includes ensuring that the information infrastructure, technological affordability and information policies are in place to sustain the development for the long haul. Factors such as overall information environment, social justice, ethics, and participation of the locals at the grassroots level are among other factors contributing to sustainability. Being at the juncture of recent policy changes, the recent expansion of infrastructure, and the rapid growth of mobile use in India contributes to the immense timeliness for raising awareness of the barriers faced by rural micro-entrepreneurs for their financial access and inclusion in e-commerce. Before delving into economic development for micro-entrepreneurs, it is important to understand the transformation of the information environment in India.

### 1.1.1 Affordable network services and devices

Mobile phone use has proliferated among lower socioeconomic populations due to the affordability of network services (Wasan & Jain, 2017). Paradoxically, rural communities that never had landlines now have access to mobile phones, skipping the conventional technological development trajectory. These communities lack basic amenities such as clean water and sanitation yet have access to mobile phones. For many rural users, mobile phones are the only way of accessing the internet. In 2017, India's leading cell phone provider, Reliance Jio launched 4G LTE networks offering practically unlimited data plans for about CAD 6 a month, making it one of the lowest data tariffs in the world (NDTV, 2017). The entry of Jio into the market forced competing firms to reduce their rates, causing India to jump from 150th to 1st in total mobile data usage (Economic Times, 2017). This is the fastest penetration by any mobile network operator anywhere in the world. Moreover, for over a decade non-branded handset have been popular with the lower socio-economic masses. These non-branded phones are reliable yet lower in cost while also supporting all the features as other smartphones (Jeffrey & Doron, 2013). They are significantly cheaper than the name brands yet claim to have many of the same multimedia functions. Hence, according to Wasan and Jain (2017) for the rural masses, mobiles have become commonplace through affordable devices and network services.

### 1.1.2 Digital India Programme

The transformation of the information environment in India is also influenced by recent public policies. In 2015, the government launched the Digital India programme that contributes to infrastructural development in rural regions (Digital India, 2017). The programme aimed to bridge the digital divide between rural and urban areas by providing access to high-speed broadband, WiFi hotspots, and digital literacy to deliver public services digitally. With the slogan for Digital India being "Faceless, Paperless, Cashless" (Digital India, 2017) the aim is to facilitate e-governance and promote financial inclusion by creating awareness for e-banking and digital payments.

### 1.1.3 Demonetization

With the Digital India programme contributing towards digital empowerment for the citizens, the subsequent goal of the government was to foster a less-cash society. In November 2016, the government demonetized all ₹500 and ₹1000 banknotes for various reasons, such as to curb the black money in the economy, to promote digital payments and more. Indian banknote demonetization came across as an unexpected economic policy shock that forced a shift to digital payments (Singh & Singh, 2017). Cashless India's agenda came as a by-product of this policy change, leading to an exponential rise in e-banking, plastic money, mobile wallets, and e-commerce (Shirley, 2017). A large number of micro-entrepreneurs, even street vendors, now accept digital payments, prompting the public to learn to transact the cashless way at a faster pace with the help of the available modes of digital payment. According to Sheetal et al. [2018], Kumar and Puttana [2018], overall a sharp jump in mobile banking was noticed post-demonetization. When the currency bills came back into circulation, a significant portion of the population moved back to using cash; this transition period lasted for 4-5 months allowing the population to rely on digital payments as one of the primary means of financial transactions. As per Kumar [2020], the digital payments in the micro-merchant and micro-enterprise segment create digital footprints such as transaction and credit histories, which helps them be included in the formal mainstream banking sector, get access to credit and eligibility for loans.

### 1.2 The Statement of Problem

With the required affordability of the mobile handsets and network infrastructure, access to the online marketplace for rural microentrepreneurs for a wider market outreach should be evident. The factors prohibiting outreach to the online marketplace could be a lack of tech-savviness and understanding of e-commerce platforms. However, there are factors that question a sustainable penetration of e-commerce such as a lack of access to information knowledge for microentrepreneurs. By auditing the impact of various government schemes on the ground, also the information needs and challenges of rural users, this research will illuminate the gaps that are leading to information asymmetries, and hindering sustainability by discouraging the rural microentrepreneurs from selling through e-commerce. The purpose of the paper is to find hurdles in awareness creation for rural entrepreneurs for adopting e-commerce as a sustainable business solution.

### 1.3 Research Questions

How can microentrepreneurs in rural communities utilize ICTs, especially mobile phones, to sell their goods through online marketplaces and ensure sustainable economic development? What factors might be prohibiting users from financial inclusion such as deficient information infrastructure, financial policies, lack of digital literacy and other socio-economic factors?

## 2. METHODOLOGY

This research utilizes a systematic literature review to collect qualitative data from peer-reviewed journal publications. The date range is from 2017 up until this point, which would give an account of digital payment penetration post-demonetization (demonetization implemented in November 2016). The search criteria would include the following keywords: India; rural, e-commerce; micro-entrepreneurs; artisans; handicraft merchants. This would help understand the social behaviour of the rural users by assessing digital literacy rates, affordances of the mobiles, available infrastructure and information resources. I will review the publications that scrutinize the ways the information environments have been transformed in the rural communities and e-commerce practices in the rural regions of India. The paper focuses on creating a structured literature review starting with challenges to e-commerce penetration in rural regions, and the current status of e-commerce operations with in-depth analysis of the e-commerce sites, newspaper reports, and articles.

### 2.1 Theoretical framework

The research will adopt Information, Communication, Technology for Development (ICTD) as a major theoretical framework. As articulated by Burrell [2009], an ICTD framework would assist me to look at "human and societal relations with the technological world and specifically consider the potential for positive socioeconomic change" [pp. 84][1]. The Cube Framework is an ICTD framework that would serve as a conceptual model depicting three-dimensional interdependencies between technology, policy, and social change [2]. As per Martin Hilbert [2012], the cube framework is based on the Schumpeterian notion of creative destruction, which "modernizes the modus operandi of society as a whole, including its economic, social, cultural and political organization" [pp. 244][2]. The cube framework would help comprehend the process of development for rural micro-entrepreneurs through mobile phones with various co-determinants at play helping determine the information dissemination gap. The cube framework would help scrutinize factors contributing to the social inclusion and/or exclusion of people in rural India.

## 3. LITERATURE REVIEW

### 3.1 Challenges to e-commerce penetration in rural India

As per a study by The Boston Consulting Group, 25% of the rural consumers find e-commerce sites and apps are hard to use. Anooja (2015) and Goswami (2016]) have analyzed the impact of the Digital India programme on the ground and have identified high illiteracy and lack of tech-savviness as gaps that restrict the effective utilization of e-commerce technologies. The discrepancies in the urban and rural literacy rates, digital infrastructure, and accessibility to information lead to a lack of access to knowledge for rural communities. UNESCO defines literacy as "the uses people make of it as a means of communication and expression, through a variety of media." In the article, "Design Studies for a financial management system for micro-credit groups in rural India" Parikh, Ghosh and Chavan (2003) claim that as per UNESCO there is 43% illiteracy in India, and "in some of the grossly underdeveloped states of northern India, illiteracy rates can hover as high as 70 to 80 percent of the population" (p. 16). This is critical in understanding the lack of literacy in some regions that might refrain the people from utilizing the available technology to its maximum potential. According to Bakshi (2019), lack of literacy is also a hurdle for professionally handling customer support. Moreover, the fear of insecure online transactions inhibits rural microentrepreneurs from using e-commerce applications; these challenges arise due to poor education, and a dearth of financial and social support (Goswami & Dutta, 2016). Despite the availability of mobile phones, digital knowledge management lags due to inefficient information use in rural communities. The lack of access to information knowledge still exists.

Furthermore, there is poor geographical access to facilitate sustainable e-commerce, which demands smooth access between the urban and the rural and remote rural regions are victims of poor transport infrastructure. A geographical challenge in the rural area is shipping goods to and from remote areas with poor access and transportation. According to Khatri, transportation and accessibility, "infrastructural problems negatively affect the productivity and profitability of the MSME sector" (p. 11) because it impacts supply chain efficiencies. As per Karnik (2016), India Post, Department of Post by the government, which is a fading away service, in the past couple of years gained a boost by collaborating with around 400 e-commerce websites, including Amazon and Flipkart.

### 3.2 E-commerce platforms to encourage participation for micro-entrepreneurs

#### 3.2.1 Bottom-up approach to the empowerment of the rural micro-entrepreneurs

In the article, "Local, Sustainable, Small-Scale Cellular Networks," Heimerl, Hassan, Ali, Brewer and Parikh (2013) propose a model to build bottom-up where the local entrepreneurs would own and operate their services for the local population. This was proposed in the context of small-scale cellular networks in Papua, Indonesia; however, this can be re-modeled for the e-commerce in the India. Also called 'inverse infrastructure,' the idea was initially coined by Egyedi and Mehos (2012) that propagates benefits of decentralization such as lower operational costs. This also creates opportunities for the locals and brings in certain freedoms (Sen, 2001). Heimerl et al. (2013) investigate the financial sustainability by scrutinizing expenses, revenues, and profitability.

#### 3.2.2 E-commerce platforms in India promoting rural artisans

The government has been ensuring the participation of the rural micro-entrepreneurs on digital channels to reach out to the wider market. This can be through the existing e-commerce platforms that leverage digital distribution infrastructure, can be facilitated through government portals or can be done autonomously by the micro-entrepreneurs through social media.

Uttar Pradesh, one of the most populous states in India, launched a program called, 'One District One Product,' (ODOP) programme, under the Ministry of MSME to encourage indigenous craft products from 75 districts across the state promoting one product that a particular district specializes in producing. As a part of ODOP, Amazon India has signed a Memorandum of Understanding (MoU) with the U.P. government to support MSMEs [Ismat, 2020]. ODOP initiative complies with the 'Atmanirbhar Bharat' project promoting the production and distribution of indigenous products. As a part of the MoU, "Amazon will provide these entrepreneurs training, account management guidance, marketing tools and world-class infrastructure of storage and delivery network to aid their progress through online selling" (Naik, 2019). Two Amazon programmes that promote rural artisans across the nation are 'Kala Haat' translating to 'Art Shoppe' and 'Amazon Karigar' translating to 'Amazon Artisans' constitutes a niche, which sells authentic crafts by Indian artisans.

Flipkart is another giant e-tailer, an affiliate of Walmart. Similar to the Amazon model in the state of Uttar Pradesh, Flipkart has signed a MoU with the state of Karnataka to promote local art, craft and handloom (Economic Times, 2020). Flipkart called this niche, 'Samarth,' which translates to 'Capable.' According to Flipkart, the Samarth programme seeks to break entry barriers for artisans by extending time-bound incubation support, which includes benefits in the form of onboarding, free cataloguing, marketing, account management, business insights and warehousing support (Economic Times, 2020). Another major e-tailer is JioMart, with Reliance as a parent company, and affiliated with Facebook, which is also WhatsApp's parent brand. Facebook has also collaborated with Reliance, which owns multiple businesses that includes Reliance Mall. Reliance is also the parent brand for Jio mobile that provided the lowest recorded prices on mobile data. According to Anand and Phariyal (2020), "Reliance-Facebook combination represents a Goliath-like opponent, especially given Reliance's track record in decimating rivals when it entered the telecoms market with Jio Infocomm and cut-throat pricing." It is interesting to observe the market dominance of Reliance and Jio, with JioMart as e-commerce, with WhatsApp and Facebook parent social media channels.

The government is planning to set up an e-commerce portal called 'Bharat Craft' for MSMEs with the State Bank of India (SBI) announced by the Chairman of SBI, Rajnish Kumar (Economic Times, 2020). Bharat Craft's initiative gears toward eliminating the involvement of middlemen and would follow a model similar to that of China's giant, Alibaba. Moreover, amid COVID-19, the Khadi and Village Industries Commission (KVIC), under the Ministry of MSME, urged e-commerce giants to support khadi mask makers. However, due to onboarding hurdles, the link-up did not materialize. In this situation, the KVIC inaugurated an e-commerce portal on their official site, which gained tremendous popularity. Upon its success, various giant e-commerce wanted to sell Khadi under their name brand. However, the government has trademarked Khadi under KVIC and denied the rights to other e-commerce portals to advertise or sell khadi (KVIC, 2020). KVIC primarily sells khadi, which stands for hand-woven cloth including khadi masks, fabrics, food items, soaps, footwear and more.

Surpassing the formal e-commerce infrastructures, various rural sellers are opting to reach out to the market autonomously through social media. In the article, "Infrastructure as creative action: Online buying, selling, and delivery in Phnom Penh," Jack, Chen and Jackson (2017) state that "the online buying ecosystem takes the rules of the local context as self-evident (i.e., the normal way of working) and "central" to the workings of the ecosystem (p. 6513). This is crucial in understanding the informal ways sellers attempt to sustain themselves in the online marketplaces. To support selling on social media platforms such as 'Dukaan' enable local shops to create an account, add items to the catalogue, set product pricing, delivery fee and sell on social media such as WhatsApp, Facebook, Instagram. This platform would assist those micro-enterprises to reach the market directly through various social media channels.

### 3.2.3 China's e-commerce giant, Alibaba's rural expansion program as a role model

China's e-commerce giant, Alibaba, provides a model for how other e-commerce platforms can strategize the utilization of digital channels to extend e-commerce to rural markets in a large and diverse developing country. Alibaba's rural expansion program provides rural service centers in the countryside that help entrepreneurs learn to sell online. These connected villages are called 'Taobao villages' and have rural retailers who manage online orders (Jain & Sanghi, 2016). The sellers from the villages receive tech support from these service centers. 'Taobao villages' act as a hub where e-commerce orders are delivered, instead of home-deliveries where rural users can pick up their orders (Jain & Sanghi, 2016). These rural service centers are run by the youth of the village, who also spread awareness about online shopping among those villagers who are unaware or reluctant towards it. Alibaba's service centers in the countryside exemplify business to customer (B2C) support model.

## 4. FINDINGS

Keeping the Cube framework in mind, in this research 'technology' refers to financial access and inclusion with the help of ICTs and infrastructure. The technology entails both hardware that is infrastructure with network towers and mobile devices and software such as app platforms on the devices. A survey of the Digital India programme and demonetization demonstrate that these recent public policies contribute to the expansion of infrastructure, the rapid growth of mobile usage in rural India and awareness of cashless payments contribute to the basic technology layer of the cube required for e-commerce to function. To address the poor transportation infrastructure for product deliveries in remote regions, India Post comes as a rescue with widespread networks across the nation to even the remotest locations.

The other layer of the cube focuses on human components, their capabilities and skills, however, lack of information knowledge is a hindrance for MSMEs in spite of available technological resources. Despite e-commerce giants signing MoUs and government reform programs in a few states, rural artisan sellers are financially excluded due to the widespread lack of awareness about e-commerce. That means a lack of freedom of choice to choose markets, lack of support in being introduced to and sustain an e-commerce environment. Human capabilities are essential to utilize technology for development to bring about social change. Social change here refers to the socio-economic sectors that are subjected to be changed with digitization. It is vital to note that semi-literacy and illiteracy are restricting users from utilizing available technology. To bring about a social change in the current scenario, human skill for the operability of e-commerce platforms is required. Lack of literacy and technical support dovetails various issues such as restricting sellers from providing professional customer support and increases fear of digital payment transactions. Another layer of the cube delves into the necessary policy instruments that fulfill the purpose of social change. Here the purpose of the social change is a sustainable development of e-commerce for rural artisan sellers. To get rural users on board there should be policies on providing the interface content and support in their vernacular and regional languages. While national languages are English and Hindi in India, each state has its regional language with a distinct written and spoken format. Currently, Amazon seller support services provide tutorials in English, Hindi and Tamil only (Varshney, 2020). Hence, there should be policies on extending the range of regional languages that would give rural artisans control and command over the interface and overall foster the e-commerce sector. The intervening guiding policies are prescribed for the government and the e-commerce giants to facilitate e-commerce in various ways such as raising awareness, providing onboarding and technical support to rural users. These policies are supposed to address the needs of the human components such as lack of tech-savviness, literacy and more.

Based on the survey of the existing e-commerce in India, I recognized 3-tiers for a sustainable e-commerce development for the rural user-base: The first tier is rural sellers collaborating with e-commerce giants such as Amazon, Flipkart, Jio Reliance-Mart and others. Collaboration with these giants means the rural users would be receiving onboarding training and technical support with an associated commission fee. The second tier is government initiative and artisan cooperatives such as KVIC and 'Bharat Craft' selling directly through their platforms, which might have less or negligible commission fee due to associated government funding and subsidies. Moreover, it is interesting to investigate informal ways sellers adapt to the online ecosystem which would also save them middlemen and other costs to place the products in the appropriate markets. Hence, the third tier is rural sellers arranging sales autonomously through social media which includes marketing and selling products on Facebook, also through Facebook and Instagram live, sending product pictures to individual clients through WhatsApp or creating WhatsApp groups for all the clients.

The third tier is a perfect example of sellers utilizing localized resources, adapting to utilize social media affordances and exercising their freedom to expand outreach to the wider markets.

## 5. DISCUSSION, STRATEGIES AND SOLUTIONS

### 5.1 Ensuring accessible e-commerce designs for illiterate users

Semi-literacy, illiteracy and lack of information knowledge are a few factors that refrain rural microentrepreneurs from not sustaining in the online marketplace. Goetze and Strothotte (2001), Huenerfauth (2002), Parikh, Ghosh and Chavan (2003) have proposed the usage of graphics within the design to facilitate usage among illiterate users. It is important to understand that e-commerce was initially originated in the Global North, and Global South adopted the e-commerce platforms as it is with the least number of changes to its basic structure and user interface. An example of this is Amazon, whose interface is the same as amazon.com and amazon.in (India). While the users from Global North and urban India would have comparable literacy to comprehend the e-commerce design principles for those platforms, the users from rural India are socially excluded because those platforms do not match their level of comprehension. Hence, for the financial inclusion of rural users, e-commerce platforms should be designed to be more accessible. In the article, "Text-Free User Interfaces for Illiterate and Semi-Literate Users," Medhi, Sagar, and Toyama (2006) state that useful platforms for illiterate users should consist of a user interface that novice illiterate users can operate with no guidance or training from anyone. Medhi et al. strongly believe that "if the UI were designed well, users would not require formal literacy, computer skills, or any external assistance in order to operate the application" (p. 72). An on-ground paper prototype experiment seeking to develop an accessible user interface for rural semi-literate and illiterate users by Parikh, Ghosh and Chavan (2003) revealed that most users could recognize the numeric keypad, its purpose, and various icons (p. 17). The users demonstrated strength in associating ideas and actions with highly representational icons relating those to the idea they meant to represent (p. 19). This is critical in recognizing the ease of the users with numbers and graphic elements despite low literacy levels. The users need to be able to operate the e-commerce platforms without formal training because access to a training programme might not always be feasible considering the huge population of rural India. The graphics-oriented user interfaces could be combined with less-text with a choice of vernacular and regional languages to assist users in easy comprehension of the content.

### 5.2 Opportunities for microentrepreneurs to sell online

The rural artisans generally sell their products within the markets in their villages, or travel outside their districts, and sometimes even travel to cities to sell their products for wider outreach. The market outreach is a tedious process for the sellers due to travel and accommodation costs among other hurdles. Additionally, the COVID-19 situation came across as an environmental intervention encouraging MSMEs to sell online. According to Gopal Pillai, VP of Seller Amazon, COVID-19 has increased customers' demand and seller registration to join online journeys (Naik, 2019). Flipkart adds in a statement that the spread of COVID-19 has urged sellers to re-think their usual mode of operating and realize the value of e-commerce, which extends their outreach to the market amid guidelines to stay away from crowded markets (Abrar, 2020). E-commerce provides the rural artisans with an opportunity to start a new business or to promote an existing one with low investment cost since stocking products in a brick-and-mortar shop is not required, which saves them rent or cost for owning a shop. It provides a merchant autonomy to sell the products based on their production speed. Mainstreaming of e-commerce among rural micro-entrepreneurs can bring a social change by reducing the inefficiencies arising due to shortcomings of the traditional channel intermediaries such as having middlemen buy in bulk for a much cheaper price and sell at high rates in cities. Switching to an online marketplace can eliminate the middlemen, which can help artisans set higher pricing for the same item being sold for a cheaper price earlier. It provides easy access to markets nationwide and also an opportunity to export internationally in some cases. According to a recent news report in India's leading newspaper, The Time of India (2020), the sellers registered for Amazon India's 'Global Selling' program noticed a 76% hike in sales for Black Friday. Hence, the adoption of e-commerce contributes to the financial inclusion of rural micro-entrepreneurs into the mainstream practice and supports poverty alleviation providing more autonomy to the sellers to pick their markets.

As per the 3-tier e-commerce resources explored earlier, it is important to note that amid high commission fees and other onboarding restrictions for rural artisans by giants such as Amazon and Flipkart, the state governments are seeking their support in promoting MSME government initiatives. The government is also promoting their independent initiatives to support rural artisans when it becomes too difficult to get assistance from the e-commerce joints, such as the KVIC Khadi sales portal. To combat the problem of poor transportation infrastructure, KVIC has tied up with India Post for item delivery and provides authority for e-tailers to set their pricing. This eliminates or reduces the commission to the e-commerce giants. This case exemplifies the power of the government in demonstrating solidarity and regulating fair-trade for the rural sellers.

Furthermore, various studies have observed the handloom and handicraft merchants autonomously selling their products by utilizing live streaming affordances of Facebook, and Instagram and sharing product catalogues through WhatsApp, in an attempt to adapt to the available resources without training. However, relying on social media especially for those who are not tech-savvy and from lower-socio economic background might bring their own set of hurdles, such as, transporting the product especially to and from geographically remote areas and digital payments. Jack, Chen and Jackson (2017) share a case of a remote region Phnom Penh in Cambodia that explains how a small firm posts the product on Facebook, the price negotiation takes place in the direct messaging service, once the purchaser agrees on the price, the delivery is coordinated with the purchaser. This tedious process involves coordination between various people, which would be difficult for rural sellers. The availability of the technical, infrastructural and policy interventions would assist the rural handicraft merchants with capacity building and generating livelihood opportunities independently and can be scaled to rural regions across the nation.

### 5.3    State Policies to provide sustainable environment for micro-entrepreneurs

The policy intervention is required from the central and state governments to further facilitate the prevalence of e-commerce among artisan sellers in rural India. Firstly, the individual states in India signing MoUs with giant e-commerce such as Amazon or Flipkart model has been working well. Hence, the government can incentivize the giant e-commerce firms to promote local artisans and help them onboard. Gopal Pillai, VP at Seller Services, Amazon India, explains that these initiatives offer help to MSMEs with required technical and operational skills, along with considering adding helpful features for those who sell through mobile phones (Naik, 2019). Secondly, the bottom-up model by Heimerl et al. (2013) can involve the local people more with running the e-commerce by leveraging local information and transportation infrastructure, and labour. Thirdly, Alibaba's rural expansion model with rural service centers helping entrepreneurs learn to sell online is exemplary. This can be scaled with the help of the government policies for the e-commerce giants to have a rural service center in villages, which would help rural sellers onboard, and support dealing with customers post-selling. The hurdles for the sellers such as return on merchandise can increase the business operation cost, while these could not be eliminated, having support can make it easy for the sellers to understand and operate within an online marketplace environment. They can receive support on appropriately fulfilling the orders, provide professional customer support and exercising control over the selling procedure. For the rural sellers "lack of on-ground presence, including poor after-sales services, is a quick way to lose customers" (Bahree, 2018). Also, this model can help overcome the transportation infrastructure difficulty by shipping the goods to a district or village at the center, where the rural population can arrange a pick-up. This complies with the 'creative infrastructural action' defined by Jack, Chen and Jackson (2017) as the resourceful and imaginative development of a homegrown infrastructure to support the ecosystem for the remote population to enjoy the conveniences of online selling (p. 6512, 6519). This also means working through the drawbacks of poor transportation infrastructure in the rural and coming up with makeshift arrangements to utilize e-commerce in the best manner. According to Kalambe (2019), the area that requires improvement includes mentorship programs and the development of a skilled digital workforce. The government has already announced the inception of 'Bharat Craft' as a platform for artisans across the nation to sell the products online, which would follow a model similar to China's Alibaba. The concept of rural service centers can be added to the structure of 'Bharat Craft.'

The key solutions for sustainable economic development are the availability of an accessible user interface for semi-literate and illiterate sellers, support to onboard and sustain e-commerce with rural service centers, policy interventions to increase the awareness of e-commerce market models. A sustainable e-commerce development would ensure poverty reduction for marginalized communities and will provide a much-needed promotional boost to their intangible cultural heritage. This research provides strategies for stakeholders to encourage rural micro-entrepreneurs to sell online and contribute to the financial inclusion of rural sellers in other nations in both the Global North and South.

## REFERENCES AND CITATIONS